# LearningCity: Knowledge generation for Smart Cities

Dimitrios Amaxilatis, Georgios Mylonas, Evangelos Theodoridis, Luis Diez and Katerina Deligiannidou

**Abstract** Although we have reached new levels in smart city installations and systems, efforts so far have focused on providing diverse sources of data to smart city services consumers, while neglecting to provide ways to simplify making good use of them. In this context, one first step that will bring added value to smart cities is knowledge creation in smart cities through anomaly detection and data annotation, supported in both an automated and a crowdsourced manner. We present here LearningCity, our solution that has been validated over an existing smart city deployment in Santander, and the OrganiCity experimentation-as-a-service ecosystem. We discuss key challenges along with characteristic use cases, and report on our design and implementation, together with some preliminary results derived from combining large smart city datasets with machine learning.

**Keywords**: smart city, data annotation, anomaly detection, machine learning, evaluation, IoT, OrganiCity.

## 1 Introduction

Smart cities have slowly been turning from a vision of the future to a tangible item, through the efforts of numerous research projects, technological start-ups and enterprises, combined with the recent advancements in informatics and communications. It is currently a very active field research-wise, with a lot of work dedicated to developing prototype applications and integrating existing systems, in order to make this move from a vision to reality.

---

D. Amaxilatis, G. Mylonas (corresponding author), E. Theodoridis, K. Deligiannidou
Computer Technology Institute and Press "Diophantus", Patras, Greece, e-mail: mylonasg@cti.gr

Luis Diez
University of Cantabria, Santander, Spain, e-mail: ldiez@tlmat.unican.es

While there is still a wealth of ongoing activity in the field, in terms of technologies competing as candidates for mainstream adoption, at least we have a number of applications slowly emerging and taking shape inside smart city instances. For example, much buzz surrounds the smart city IoT testbed and experimentation concept, like in the case of SmartSantander [29]. Another example is the utilization of open data portals in smart cities, like CKAN [2], an open source solution provided by a worldwide community, and Socrata [5], an enterprise solution backed by an IT company. Additionally, protocols like MQTT and communication technologies like LoRa or NB-IoT are used in recent smart city research projects to provide real-time communication with the deployed infrastructure, and progress towards becoming Internet standards.

However, essential answers are to be found revolving around a central question: what do we do with all of these data collected, and how do we make sense out of them by extracting knowledge, i.e., something actually useful, going beyond a technology demonstrator? Related to the previous question, we also need to find a way to provide usefulness to citizens, involving them in the smart city lifecycle, and engaging them in the city shaping. By creating more "useful" information out of raw sensors, or other kind of data representing observations of the urban environment, better ways to discover data streams and easy ways to utilize the information will create significant acceleration in the smart city ecosystems. For example, certain events generate data reported by the city sensing infrastructure, but are, more often than not, missing an appropriate description. Consider the case of a traffic jam inside the city center. It generates sensed values in terms of vehicles speed, noise, and gas concentration. In addition, in most cases, multiple devices or services, while missing useful correlations in the data streams, report such values.

We believe that adding data annotations to smart city data, through machine learning or crowdsourcing mechanisms, can help in revealing a huge hidden potential in smart cities. In this sense, one key aspect is how to traverse through this sea of smart city data in order to decide where and what to look for before adding such annotations. In addition, we also have to analyze correlations from findings after having processed this data, in order to uncover the hidden information inside them.

In this work, we discuss the design and implementation of JAMAiCA (Jubatus Api MAChine Annotation), a system for aiding smart city data annotation through classification and anomaly detection. On the one hand, it aims to simplify the creation of more automated forms of knowledge from data streams, while on the other hand it serves as a substrate for crowdsourcing data annotations via a large community of contributors that participate in the knowledge creation process. We strongly believe that communities like data scientists, decision makers and citizens should get involved in deployments of future Internet systems, for them to be practical and useful.

In order to validate our approach, in this work We include a number of illustrative use cases for which we provide some evaluation results. These use cases utilize a combination of data sources that provide insights to the actual conditions in the center of the city of Santander, like parking spots, traffic intensity and weather sensors, along with the load in wireless telecommunication networks. Based on our preliminary analysis, our findings show some interesting correlations between the aforementioned datasets, that could be of interest to city planners, local authorities and citizen groups.

Moreover, the system present here was designed, implemented, employed and evaluated inside the context of OrganiCity[1] ecosystem. OrganiCity, as a smart city technology ecosystem, aims to engage people in the development of future smart cities, bringing together three European cities: Aarhus (Denmark), London (UK) and Santander (Spain). Co-creation with citizens is its fundamental principle, i.e., defining novel scenarios for more people-centric applications inside smart cities, exploiting the IoT technologies, heterogeneous data sources, and using enablers for urban service creation and IoT technologies. Fig. 1 provides an overview of OrganiCity. In short, the project aims to provide an Experimentation-as-a-Service (EaaS) platform [10], i.e., it is designed to make data streams from diverse sources inside a smart city available to various "consumers", like IoT experimenters, SMEs, municipalities, etc. At the same time, it aims to enable the participatory engagement of communities in co-creating urban knowledge. This is done by means of end-user applications that provide meaningful representations of the produced smart city data, and "tools" that will allow these end-users to make their own contributions.

Regarding the structure of this work, we first report on previous related work, and continue with a discussion on challenges associated with knowledge creation in smart cities. We then present a small set of use-cases to highlight how our system relates to this vision. We continue with a presentation of our design and system architecture, complemented with a description of our current implementation and some preliminary results we have produced so far. Finally, we summarize the main contributions brought about by the JAMAiCA system, and highlight some aspects that will be tackled by exploiting the annotation system.

## 2 Previous Work

Although there have been a number of recent studies and applications aiming to combine human and machine intelligence, research in this field is still at its infancy stage. In [17], authors present a vision on the potential of combining patterns of human and machine intelligence, identifying three possible patterns sequential, parallel and interactive. Moreover, in [11] authors present a crowd-programming platform that integrates machine and human based computations. Their system integrates mechanisms for challenging tasks like human task scheduling, quality control, latency due to human behavior, etc.

---

[1] Co-creating digital solutions to city challenges, https://organicity.eu/

A key element in most of the approaches is the use of a taxonomy or ontology, which are ubiquitous in organizing information. They represent formal structures to represent entities, to organize entities to categories, to express their relations and to map data objects to abstract concepts expressing meanings, entities, events etc.

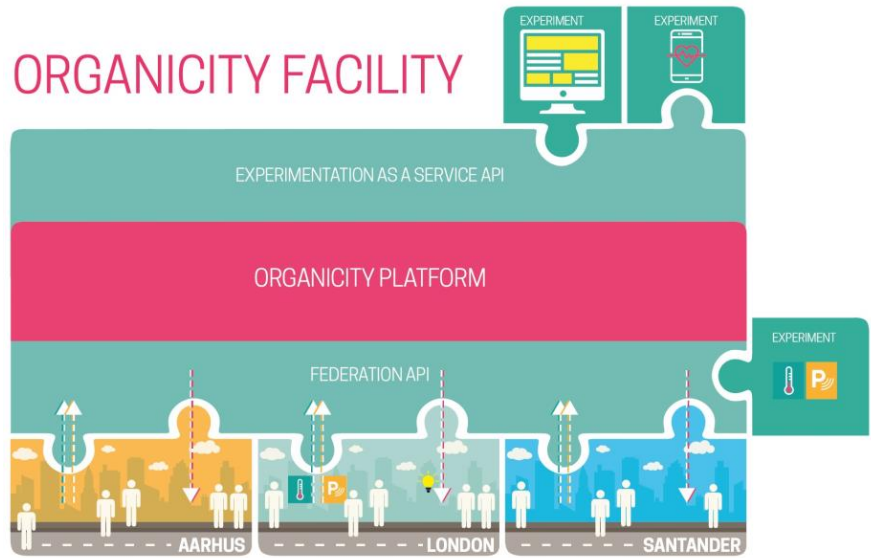

**Fig. 1** A high-level view of OrganiCity as a smart city platform. The platform offers a Federation API through which 3 European cities (Santander, Aarhus, London) with smart city infrastructure are "federated". This federation enables experimenters that utilize the experimentation-as-a-service API of the platform to see these 3 cities through a single interface. Additional options are available through smartphone experimentation and additional cities' federation with the platform.

Most of the modern social networking applications and online collaborative tools are heavily relying on an underlying taxonomy. Building and curating a taxonomy is a challenging task that requires deep knowledge of a specific domain and the corresponding data characteristics, usually performed by a small group of experts of the application domain. In contrast to proprietary domains, folksonomies are quite popular in online applications and they organically created by their users. Such taxonomies usually have weaknesses like double entries, misclassified tags, entries with typos or ambiguities in the classes, and so on.

In [14], the authors propose a workflow that creates a taxonomy from collective efforts of crowd workers. In particular, the taxonomization task breaks down into manageable units of work and an algorithm coordinates the work that involves mapping to categories, identifying the best mappings and judge the relevance of the associated categories. Although there exist taxonomies and tagging of objects with keywords of the taxonomy, the problem is that there is no common agreement about the semantics of a tagging, and thus every system uses a different representation. In [21], an effort for the development of a common tagging ontology with semantic web technologies is described.

Designing and developing smart cities is a concept that has drawn tremendous attention from the public and the private sector. Each one of the scientific disciplines like urban engineering, computer science, sociology and economics, provide unique perspectives on making cities more efficient. In most of these cases, multidisciplinary approaches are required to tackle complex problems. Projects employing machine and crowdsourced learning techniques started to take shape the last years.

SONYC [12] is an example of a project with a very well-defined use-case, employing machine-learning algorithms to classify acoustic readings into various types of noise encountered inside an urban environment. It is a very interesting approach, with similarities to our vision of providing a generic substrate to simplify the process of knowledge extraction and data annotation contributions. Moreover, learning from the crowds, by using the crowdsourced labels in supervised learning tasks in a reliable and meaningful way is investigated in [28, 33].

A large number of projects are trying to leverage modern information and communication technologies, like IoT/Future Internet and the semantic web, in order to build novel smart city services and applications. An example is the SmartSantander project [29], which has developed one of the largest Future Internet infrastructures globally, located at the center of the city of Santander in Spain. A well-established citywide IoT experimentation platform that moved testbeds from labs to the real world and that offers experimentation functionality, both with static and mobile deployed IoT devices, together with smartphones of volunteers inside the urban areas. [22] discusses the SOCIOTAL EU project, which attempted to tackle co-creation aspects inside smart cities. Another example is CitySDK [1] that tries to harmonize APIs across cities and provide guidelines about how information should be modeled, propose ways to exchange data, and how services and applications should be designed and developed. The project benefits from semantic web technologies, and focuses on application domains like citizen participation, mobility and tourism.

CityPulse [26] introduces a framework for real-time semantic annotation of streaming IoT and social media data to support dynamic integration into the Web. The framework employs a knowledge-based approach for the representation of the data streams. It also introduces a lightweight semantic model to represent IoT data streams, built on top of well-known models, such as TimeLine Ontology, PROV-O, SSN and Event Ontology. In terms of creating high-level concepts from the large amount of data produced, another similar approach has been carried out in [16]. The latter approach introduces a methodology to automatically create a semantic ontology, without requiring preliminary training data, using an extended k-means clustering method and applying a statistical model to extract and link relevant concepts from the raw sensor data. In [13] the authors propose principles for semantic modelling of city data while in [27] authors propose a technique to extract hidden structures and relations between multiple IoT data streams. The method employs latent Dirichlet allocation (LDA) on top of meaningful abstractions that describe the numerical data in human understandable terms.

In [18] authors present a IoT deployment enhanced with a machine learning and semantic reasoning layers on top. Moreover, in [25] they explore by surveying the application of Deep Learning (DL) techniques on IoT and smart city streams and try to surface challenges, limitations and opportunities. [9] discusses the IoT field from a data-centric perspective. PortoLivingLab [30] is a smart city deployment project supporting multi-source sensing from IoT deployments and crowdsourcing in order to achieve city-scale sensing focusing on weather, environment, public transport, and people flows. Finally, they present a set of use cases that provide key insights into the status of city of Porto, Portugal. In [24], the authors discuss smart mobility scenarios that are representative for big cities, and especially in China.

Regarding the infrastructure described and used for our evaluation, [29] introduces aspects of the utilized infrastructure, while [31] discusses issues and lessons from the deployment and operation of such a large IoT smart city infrastructure. In practice, through our evaluation we have detected issues in the operation and continuity of data, which as mentioned in [31] are aspects which could be serviced by systems like the one discussed here. The use case of parking inside the city is presented in detail in [23]. The issue of overall data quality in IoT is discussed in [20].

The work presented in this article, acts as both an end-user tool and a service for other applications to extend the data annotation functionality of a wider smart city ecosystem. Currently has been employed and tested in the OrganiCity smart city technology infrastructure and especially data stream discovery services like Urban Data Observatory (UDO)[2]. An earlier version of the work discussed here was presented in [15], while a detailed discussion of some of the findings regarding experimentation in OrganiCity is included in [10]. In the latter work, we provided a detailed discussion on the design and implementation, along with additional related results.

## 3 Data Annotation in Smart Cities - Challenges

In this section, we briefly discuss a set of key challenges regarding data annotation in smart cities. We can partition them in two fundamental objectives:

- enable a more engaging and secure experience for citizens/contributors.
- produce a more meaningful results/observations from the system side.

Privacy and overall security issues are a central challenge in the context discussed here. Consider the case of a volunteer taking noise level measurements along his daily commute or being tasked to add annotation contributions by a smart city system based on proximity to certain events. Even in such simple scenarios, anonymization techniques should be used to ensure that neither personal data, nor interactions are revealed.

---
[2] https://docs.organicity.eu/UrbanDataObservatory/

Another important issue is the correlation of different types of smart city data that can potentially point to the same event, in other words, how to facilitate knowledge extraction through such data. We currently have data produced by IoT infrastructure installed inside city centers. However, there is relatively small research focus on discovering relations between these data. For instance, we may analyze if noise level readings are related to a live concert event or can be attributed to another event produced by a specific situation (e.g., traffic jam) taking place somewhere inside the city.

Moreover, there is the issue regarding the nature of data available in smart city data repositories, being data inserted by humans or IoT infrastructures. Both sources can be unreliable, or even malicious. With respect to sensing infrastructure, we also have the issue of the hardware malfunctions, as well as spatiotemporal effects on the data produced. In most cases, the hardware utilized aims for large-scale deployments, thus being not so accurate or having calibration issues. Additionally, environmental conditions, e.g., excessive temperature or humidity, may have an effect on the sensitivity of the sensing parts. In this regard, one key aspect is how to produce data annotation based on such an infrastructure, which can function with a varying degree of credibility during a single day. Reputation mechanisms are an example of measures that can aid in this direction, either human or machine-based, in order to filter out less reliable data sources.

The issue of end-user engagement with respect to data annotation and knowledge extraction is, in our opinion, another major challenge. We also think that user contribution is twofold: end-users can contribute to a smart city system by adding data annotations, but also contribute data through incentivation or gamification. Although most current crowdsourcing platforms utilize a desktop or web interface, the crowdsourcing of data annotations does not have to be limited to that. It can be also performed through smartphones and be incorporated to the user's everyday life. The interaction of end-users through such a tool could help in relating in a more personal way and maintaining the interest in participating. Moreover, annotation of events or sensed results could be more interactive and focus at users, or even user groups, near the actual space of the event in question.

Smart city facilities usually integrate a large number of data sources of various types sharing observations for environment, air quality, traffic, transport, social events and so on. These data sources might be static (they are not streaming data and have a fixed value until they are updated by an offline process) or might be dynamic (streaming data constantly). Building a taxonomy on this multi-thematic environment is not straightforward, since some tags subcategories might be shared between different types of data sources, while others might be orthogonal. In addition, as the dynamic data sources have a temporal dimension, annotations might characterize the overall behavior, and observations, of data sources or observations falling into a specific time interval. Furthermore, as data sources might be mobile (e.g., an IoT device on a bus or a smartphone) an annotation might characterize a specific location inside the city within a specific time interval. Embedding these spatiotemporal characteristics in the taxonomy introduces new requirements and extensions to traditional methods. Standardization bodies, like W3 Web annotation data model and protocols, do not cover sufficiently these requirements.

Finally, implementing machine learning algorithms suited to smart city data and real-time processing is another major challenge. Handling citywide data introduces additional complexity, especially when considering relations between different data types and sensing devices. Current mobile devices have enough processing power to handle a broad set of use-cases, especially when dealing with data from integrated sensors (e.g., [6] uses on-device processing to classify urban noise sources). This could also be utilized as a means to enhance privacy, since processing would be performed locally, without requiring sensitive data to be uploaded to the cloud.

### 3.1 Use cases

We now proceed with a set of characteristic use cases, aiming to highlight our vision of the annotations system, and to provide insights to tackle the aforementioned challenges.

- *IoT sensors to create better running and biking routes*: this use-case utilizes mobile and smartphone/smartwatch sensors to monitor environmental parameters so as to infer better routes for running and biking in terms of healthy environmental conditions. Parameters that could be sensed include air quality, noise pollution, pollen concentration, condition of roads, etc. Machine learning techniques could be used to identify anomalies in the sensed data, such as high pollutant or particle concentrations, or locations with high noise levels. Alerts regarding such events could be sent by the system to participating end-users to quantify or validate such data through annotations. Another use of data annotation contributions could relate to the sentiments of participants for their surroundings.
- *The soundtrack of the city*: the concept is to create the aural and noise level maps of cities. This includes the use of the microphones of smartphones to record noise or distinctive sounds of the urban landscape. Participants could use data annotations to pinpoint street musicians, sounds from birds or other animals and their location, or sounds from public spaces like train, bus stations, or city halls, etc. Machine learning techniques could be used to generate general classifications that could subsequently be made more specific by end-users providing additional data annotations. Users could also add descriptions and their sentiments towards places and sounds.
- *Smart city event correlations*: a diverse smart city IoT infrastructure could "record" the same event from different aspects; a traffic jam could take place at a certain point in time (traffic data), while creating certain side effects, such as noise from car horns or engines (noise data), unusual levels of pollution (air quality data), etc. Since this kind of data is being fed to the system with similar spatiotemporal characteristics, such anomalies can be detected and correlated on a first level and then be validated by end-users to define additional correlations.

## 4 Architecture

As outlined previously, OrganiCity federates existing smart city infrastructures, integrating urban data sets and services. Federated resources are exposed, in this context, through a unified experimentation service and a central Context Broker [3]. Based on this existing architecture, our data annotation service, JAMAiCA, is designed to operate over the updates provided by the Context Broker, in order to provide additional knowledge, increasing their value and usefulness. JAMAiCA is capable of consuming, processing and annotating individual data points to produce temporal annotations or nearby measurements to generate spatial annotations.

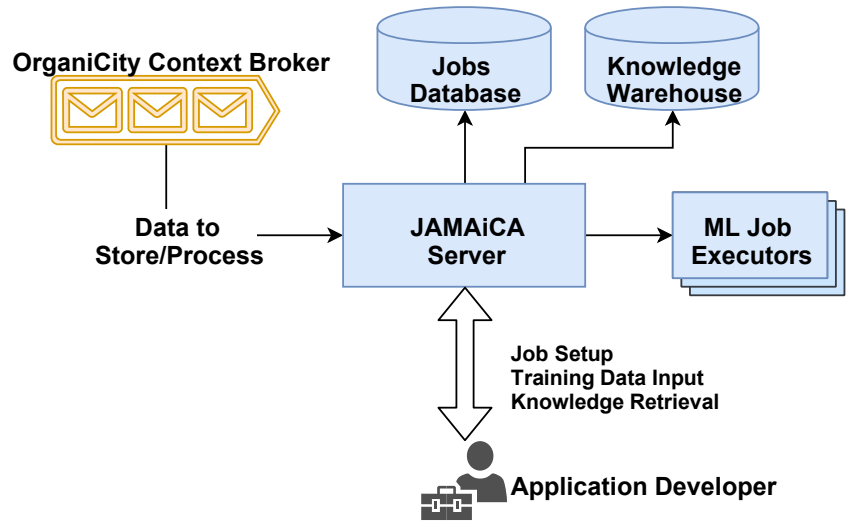

**Fig. 2** Integration of the JAMAiCA Service with the OrganiCity infrastructure. Interested application developers can utilize the datasets available from OrganiCity, by using the interfaces of the system to setup, configure and monitor the execution of machine learning jobs.

Fig. 2 presents the integration of the JAMAiCA service with the OrganiCity infrastructure. It also presents the two main software components and the other building blocks employed in order to generate and store the annotations for the IoT data received.

The core module, the JAMAiCA server, offers both an API and a web interface for users to setup, configure and monitor their machine learning jobs. The configuration of each job is stored in the Jobs Database module. Once a job is setup, the JAMAiCA server subscribes for data updates to the Context Broker and sets up a proper executor for the machine learning job. Each ML executor is configured based on the job configuration and fed with the training data provided by the user. When a new sensor measurement is received it is passed to the respective executor to be analyzed.

The Knowledge Warehouse is responsible for maintaining a directory of all possible annotations in the form of **tags**. Tags are simple indicators of the annotated parameter, similar to the way tagging is performed in photos in social networks, or the use of hashtags in social status updates. **Tag domains** are created as collections of tags (e.g., high, normal and low) with a similar contextual meaning. Tag domains can be generic as those mentioned before or more application specific (e.g., the tag "contains a traffic light" for images). Users of the system can either select one of the tag domains already available, or create their own specifically for their application. The outcomes of the analysis from the machine learning executors are also stored in the Knowledge Warehouse with additional notes that can be numeric or text values. These notes can be user comments or a value that describes the abnormality of an observation, or a confidence indicator for the classification.

The JAMAiCA server is capable of performing both anomaly detection and classification jobs over the data streams formed by the updates from the Context Broker. In both cases, after the annotation jobs are added to the system, the initial training data are submitted and the back-end is trained. Annotation begins and for each data point examined the results are stored to the Knowledge Warehouse. Regarding the machine learning executors themselves, the system is agnostic of the actual machine learning frameworks, as it is capable of using multiple external services for the job. This gives us the flexibility to experiment with various machine learning frameworks and expandability to easily provide extra functionality in the future. In our case, we evaluated during the development of the tool two distinct solutions: Jubatus and JavaML. More information on the operation of JAMAiCA is provided in the following section.

## 5 Implementation

In this section, we discuss the technologies used for the implementation of the JAMAiCA Server and Knowledge Warehouse, together with details regarding data communication and the available end-user interfaces. Regarding the implementation of the system, it is openly available on GitHub[3] via OrganiCity's repository, along with user guides and examples.

### 5.1 Communication and frameworks used

Regarding communication, the provision of data to our system is done either directly, or through an NGSI context broker [4]. Additional options like ActiveMQ or MQTT message queues can be implemented and then be added to the system. For our main use case, JAMAiCA uses a context query, provided during the creation of the machine learning job, to register for updates on an the main OrganiCity context broker. This query acts as a set of selection parameters for the devices and sensors the job is interested in. In OrganiCity, the FIWARE Orion Context Broker is used. After a subscription is established, the Context Broker uses POST HTTP requests to notify our system of the newly received data following the NGSI specification. We also offer users the option to manually send data to our system via a similar HTTP POST request to a per job auto-generated endpoint. The format of the HTTP body needs to be the formatted according to NGSI specification for the sake of simplicity.

Both the JAMAiCA server and Knowledge Warehouse, are implemented using Java and the Spring Boot framework [7]. Spring Boot is Spring's convention-overconfiguration solution for creating stand-alone, production-grade Spring-based applications, as it simplifies the bootstrapping and development stages. It eases the process of exposing components, such as REST services, independently and offers useful tools for running in production, database initialization, and environment specific configuration files and collecting metrics.

In our case, we implemented both interfaces as RESTful web services. The API of the JAMAiCA server offers methods for handling primary HTTP requests (post, get, put and delete) that correspond to CRUD (create, read, update, and delete) operations on the *jobs database*, respectively. As a result, it allows experimenters to add and manage annotation jobs through their applications. A machine learning job can be either an anomaly detection or a classification process. Since both jobs require initial training data, additional methods that allow training a machine learning instance for an existing job are available.

### 5.2 Machine Learning frameworks

In order to perform the analysis of the data, we used the Jubatus Distributed Online Machine Learning Framework [19] and JavaML [8]. In general, in the smart city plane it is usually not practical to use conventional approaches for data analysis by storing or analyzing all data as batch-processing jobs. Instead, our system processes data in online manner to achieve high throughput and low latency by using multiple frameworks and executors for online and distributed machine learning. It also processes all data in memory, and focuses on the actual operations for data analysis to update its the machine learning executors instantaneously just after receiving and analyzing the data. For each annotation process, we deploy a dedicated instance based on Jubatus or JavaML and feed it with the provided training data. Our service communicates with each instance using a wrapper with a common interface. This setup allows us to horizontally scale the machine learning infrastructure on demand. An example of the graphical user interface of the system can be seen in Fig. 3.

---
[3] https://github.com/OrganicityEu/JAMAiCA

**Fig. 3** An instance of the user interface of JAMAiCA for configuring and monitoring classification jobs. A basic classification job is set up in this case, related to light levels, where we see on the left part the description and training data for the job, while on the right there is a live feed with new values being classified.

### 5.3 Knowledge Warehouse

The Knowledge Warehouse uses Neo4j [32] to maintain the taxonomies generated by the **tags** and **tagdomains**. Neo4j is a graph database that leverages data relationships and helps us build an intelligence around the entities stored in our system and the relationships between them. By traversing the relationships between **tags** that comprise a tag domain, we can easily create suggestions for the appropriate tags a new annotation application may use. Also, the relationships between annotations can be used to extract higher knowledge for the cities or the users of the system, especially when augmented with location and time meta-data in order to identify events or situations that arise inside the cities. For example, an application can query for streets of the city where high atmospheric pollution and low vehicle speed is detected (indicating a possible traffic jam) and advise drivers to use alternative routes or means of transport, when combined with information about the local subway timetables.

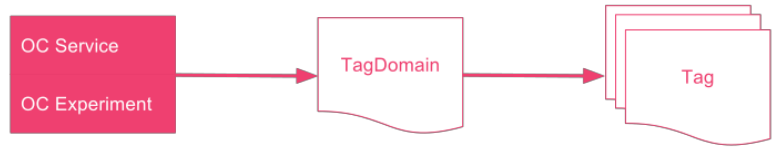

**Fig. 4** OrganiCity Services, Experiments, Tag Domains and Tags.

The underlying data model of the Knowledge Warehouse and the relations between its entities are depicted in Fig. 4:

- **Tags** represent the actual annotation labels to be used by users of the system.
- **TagDomains** represent collections of tags. Usually a tag domain is associated with a service and/or an experiment specifying which tag domains they will use.
- **Services** represent utility/urban services. An example of a service might be garbage collection, or noise monitoring. The basic usage of service entities is the organization and discovery of tag collections (e.g., what **tags** are usually used for characterizing the noise level sensors).
- *Experiments* are created by users of the system.

In addition to the entities presented above, the following concepts come to glue together the annotation schema with the OrganiCity facility entities:

- **Annotations** are relationships between the **assets** of the OrganiCity and a **tag**.
- **Assets** are entities inside the OrganiCity facility that can be annotated. The assets are not stored in the internal database of the system but referenced by the added annotations.

An example of how Assets are annotated inside the Knowledge Warehouse is presented in Fig. 5. In the figure, two Assets of OrganiCity (the red circles) are annotated with two different **tags** (the blue circles) that belong to a single **tag domain** that describes the traffic conditions in the city.

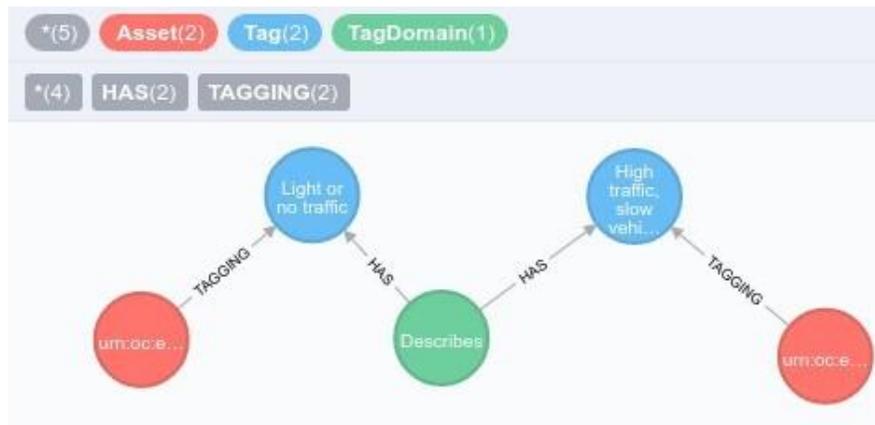

**Fig. 5** Example of Data Annotations and the Relationships created using the Neo4j server web interface.

### 5.4 End-user interfaces and integration with existing tools

Both modules presented above do not provide dedicated user interfaces for OrganiCity end users to add/validate or vote for available annotations. To handle this aspect, we provide interaction between users and JAMAiCA based on OrganiCity tools like the Urban Data Observatory (UDO [4]), a smartphone experimentation application developed for Organicity (Sensing on the Go [5]), or other applications developed by OrganiCity Experimenters themselves. The Urban Data Observatory facilitates the search among the Assets of OrganiCity, and users can use it to identify assets relevant to their interests. This is achieved by allowing for easy discovery and filtering of assets, based on geolocation, types, metadata, and even reputation or recommendations scores based on the experience, reliability and opinions of other users. The JAMAiCA information presented above is offered to users by a bidirectional interaction with annotations allowing for:

- Looking up annotations of particular assets.
- Filtering Assets based on annotations already stored in the JAMAiCA.
- Requesting to users to validate existing tags generated by JAMAiCA.
- Addition or deletion of tags for a particular asset on demand.

An example of how the annotation information is displayed on the UDO is available in Fig. 6.

---

[4] Organicity's Data Observatory, https://observatory.organicity.eu/

[5] https://play.google.com/store/apps/details?id=eu.organicity.set.app

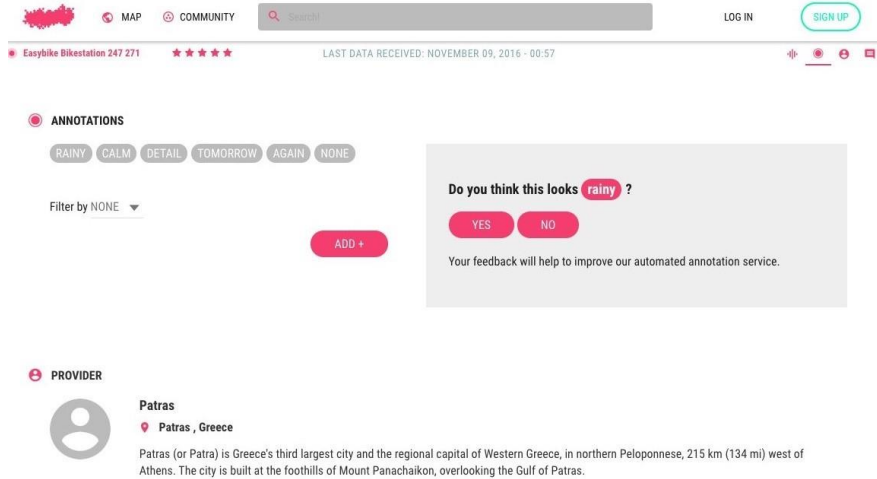

**Fig. 6** The Annotations module interface on the UDO. The end users are presented with the existing information and suggestions on how to improve data and add new aspects to the existing ones.

The JAMAiCA annotations are also integrated into the Assets Discovery Service of OrganiCity, which is the back-end that allows the UDO to perform the data filtering. This is especially important since it provides end-users with the ability to filter and search assets based on the stored tags. To provide this information on the Asset Discovery Service, annotation information is pushed to the Asset Directory Service every time they are created, updated, or deleted. However, this interface is not aimed at replacing the JAMAiCA API when retrieving the complete annotation for a specific asset, since the information stored on the Asset Discovery Service are in an aggregated and limited format.

As an additional interface for adding and collecting annotations, we have developed a specific view on the "Sensing on the Go" smartphone application that allows users to add and modify Asset annotations as part of their Experiment. This view loads a map view of the nearby area based on the user's location together with the available Assets in their vicinity. On top of these pre-existing Assets, participants of experiments can add annotations by simply selecting the appropriate **tag**. Experimenters can retrieve the generated annotations from the system and associate them with any data that their users have also collected from the phone sensors. The view for adding annotations from this application is available in Fig. 7.

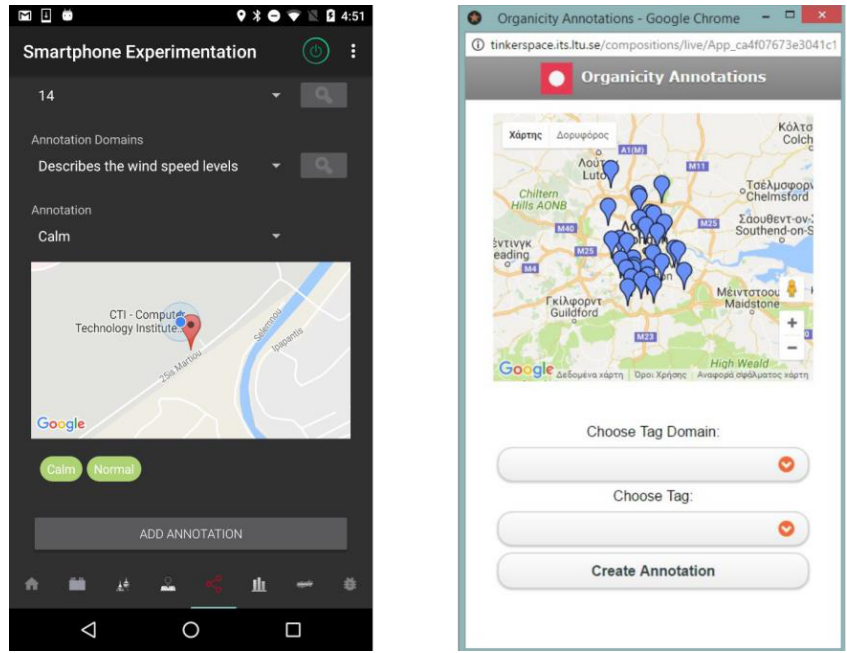

**Fig. 7** Screenshots from the smartphone interface of data annotation-related software: Sensing on the Go app (left) and Tinkerspace interface (right). On the smartphone app, and as part of an experiment running on top of OrganiCity, developers can add options for creating annotations during the experimentation process. Such annotations are added by end-users/volunteers who offer to execute such experiment on their Android smartphones, while moving through the city. On Tinkerspace, the displayed interface is an option that can be added to an application for the system, to insert annotations.

An additional end user interface is available through integration with the Tinkerspace platform. Tinkerspace [6] is an online tool for building simple smartphone "applications", that is part of the OrganiCity toolset offered to end-user communities.

A number of functionality "blocks" are available, upon which users define their own rules for processing and I/O. In order to support Data Annotation in Tinkerspace, we have provided a new version of existing blocks to support annotation. To showcase the operation of the blocks we designed, we used the visual programming interface provided by Tinkerspace to generate our own application for annotating using smartphones. The interface of the Tinkerspace application can be seen in Fig. 7.

---

[6] http://www.tinkerspace.se/

## 6 Results - Discussion

In this section, after having introduced our design and implementation, we proceed to discuss some illustrative cases that showcase both the usefulness of our system and the capacity of extracting knowledge out of smart city data. We base our evaluation on data produced by the smart city infrastructure available at Santander, Spain.

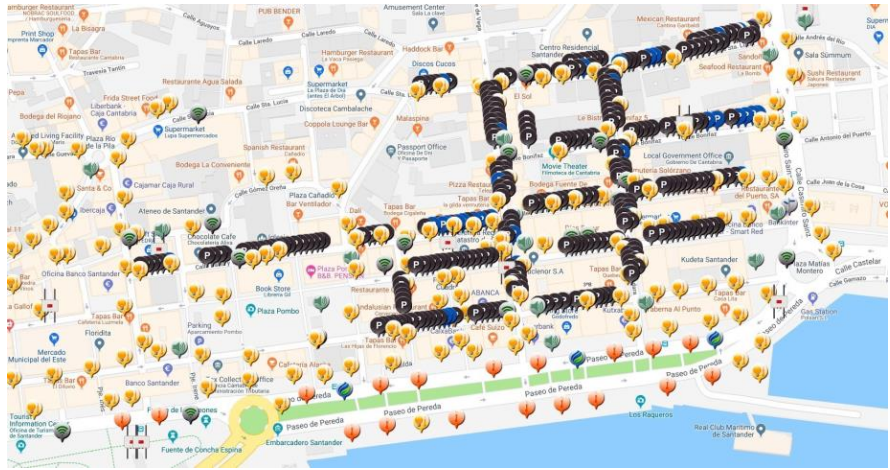

**Fig. 8** The downtown area in Santander where the majority of the IoT nodes which generate our data sets are installed. It is roughly a 800 by 400 metres area in the center of the city, containing parking, noise, weather, traffic and e-field sensors.

We have focused our analysis on the following data sources:

- Parking spots sensors: a number of sensors are installed under the actual parking spots at the centre of Santander, continuously monitoring the available parking spots.
- Traffic intensity sensors: a number of sensors are installed at big traffic junctions, mostly locate at the outer parts of the city, monitoring and counting the number of vehicles. The readings are extrapolated to 1 hour, and are measured in vehicles per hour.
- Weather sensors: weather stations are scattered around the city, producing readings that are categorize the type of weather. Values range from 0 to 11, based on the type of weather ('0' meaning sunny weather, '3' is cloudy, '7' is heavy rain, '11' is hail).
- Electric-field sensors: the e-field sensors measure the electromagnetic field in downlink communications (base-station to users) and for all operators together, for the telecommunications bands used in Europe (2G-900MHz, 2G-1800MHz, 4G-1800MHz, 3G-2100MHz). It is an indirect measure of the activity inside the city centre.

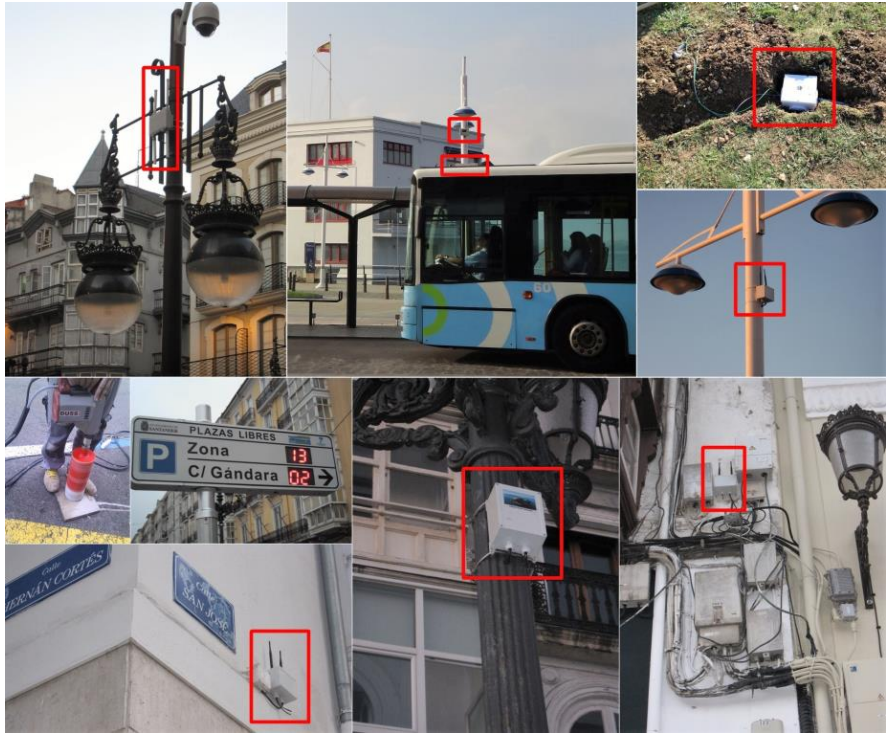

**Fig. 9** Some examples of IoT devices deployed inside the center of Santander. Both stationary and mobile IoT nodes are used, installed on lamp posts, on top of buses, buried beneath the soil in public parks, or on the walls of buildings at the streets of Santander's center.

For our evaluation purposes, we selected (see Fig. 8) a subset of the SmartSantander infrastructure[7]. The area we selected is located at the center of the city and contains the vast majority of the parking sensors in the city, apart for the other types of sensors. It is located inside the main cultural and commercial district of the city, containing lots of offices, services, banks, shops and restaurants, i.e., it is one of the busiest areas in Santander day and year-round. Examples of deployed devices are depicted in Fig. 9. Regarding the time in which the measurements were produced, data were generated during 2017, with the majority of the data examined here generated during the second half of 2017. This is due to the fact that this was the period with the smallest number of disruptions in data continuity, i.e., with less gaps in the datasets examined.

Fig. 10 provides us with a base understanding of how the parking spots in question are used by the citizens of Santander throughout the days we analyze. As we can see in the figure, the number of the available parking spots typically peeks around 3:00 in the morning and reaches the lowest points twice during each day, once around 10:00 in the morning (office hours) and around 17:00 in the afternoon when the stores are again open, and people return to their work or visit the commercial district of the city. Another important characteristic we can identify in the data of this figure is that the number of the available parking spots in the area on Saturday mornings is higher than the rest of the days in the morning (96 vs 80 available spots). Similarly, on Sunday mornings we can see that people come to the area a bit earlier than on the rest of the days (around 9:00) and occupy more parking spots (40 vs 47 available spots) almost until the evening, when the spots are again free.

---

[7] For a map view of the whole installed infrastructure, please visit: http://maps.smartsantander.eu/

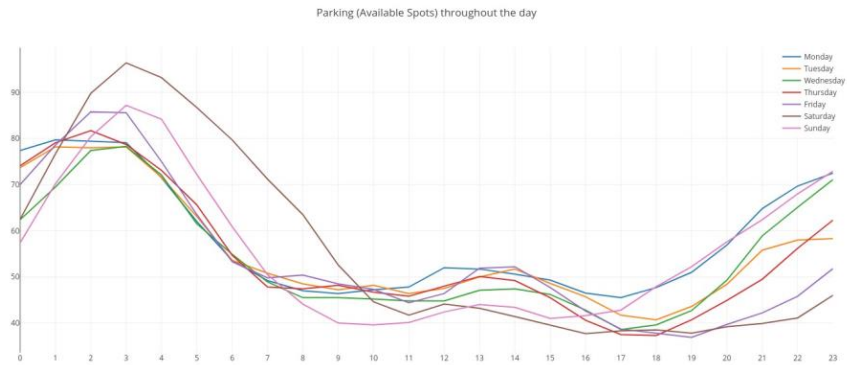

**Fig. 10** Available parking spots at the center of Santander for each day of the week between August 15 and December 31, 2017.

Fig. 11 depicts the available parking spots at the center of Santander for a given time period, while Fig. 12 depicts, in more detail, the values of e-field for a given location in the same area and available parking spots for a specific week. In many cases, there is a strong inverse correlation between the e-field values and the number of available parking spots at the center, as displayed in Fig. 13.

Regarding Fig. 14, we showcase the e-field values from 3 locations in the center. Overall, it seems that e-field values follow similar patterns, and the sensor that we consider for our study in the previous figures (*efield2*) is the one that records the higher values, i.e., it represents the busiest area. There seems to be an average e-field value, that when it starts to change rapidly, parking spots are also declining rapidly within a limited amount of time, and go from 80 to 50 available parking spots quickly. It seems that the variability of e-field could serve as an indicator for creating a prediction for when the parking will start filling up.

Furthermore, throughout the year there are in some cases anomalies in parking values and e-field values that need context, e.g., during specific days we noticed large spikes. One additional finding is that there is a large difference in availability of parking spots between summer and winter periods; in fact, values almost double in winter. This is most probably due to the fact that there are several schools and university buildings close to this area, and as soon as the main activities for the school year begin, there is a very noticeable difference in parking spots availability.

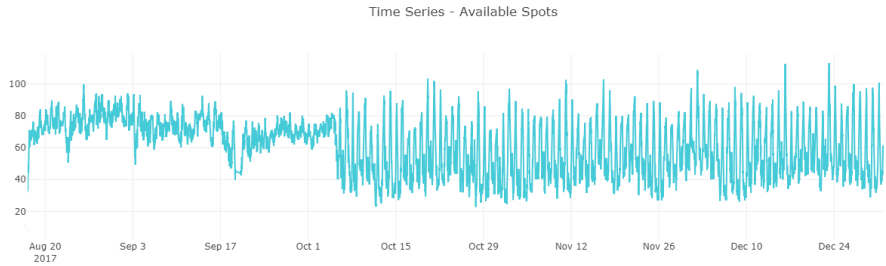

**Fig. 11** Available parking spots at the center of Santander between August 15 and December 31, 2017.

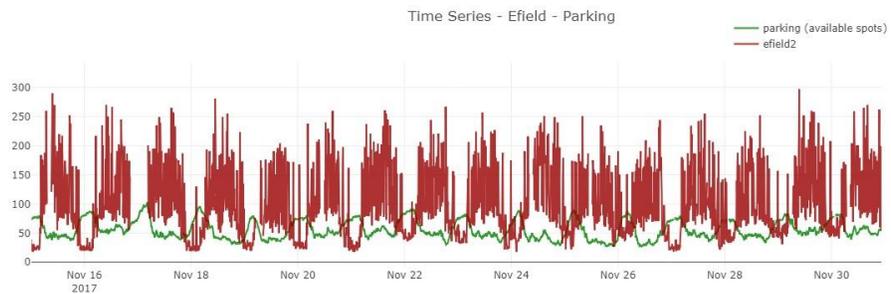

**Fig. 12** Available parking spots at the center of Santander and e-field measurements for the area containing the parking spots, between November 15 and December 1, 2017.

One other interesting finding from our graphs is that, for this particular area with its specific characteristics in mind, it is quite possible that e-field readings can, to a certain extent, substitute parking sensors. In an application scenario which is based on the overall availability of parking spaces inside the center, the e-field readings can provide a good approximation of the general picture. In the case where there are specific parking spots of a special type, e.g., parking spaces reserved for people with disabilities, then the approximation does not hold as well.

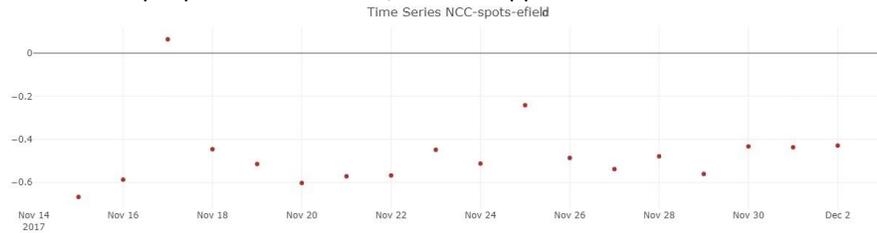

**Fig. 13** Pearson correlation per day for parking spots and e-field measurements, between November 15 and December 1, 2017.

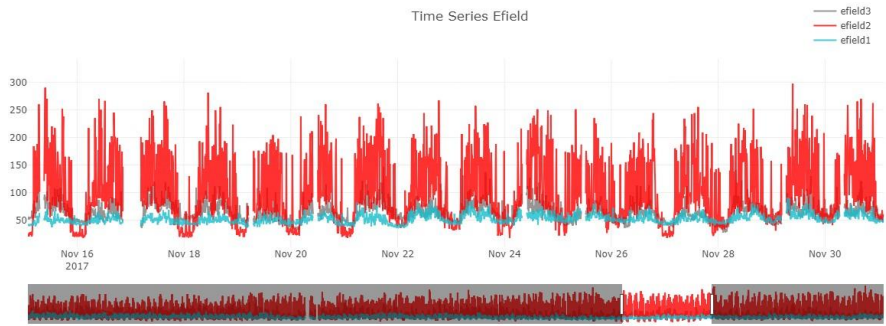

**Fig. 14** E-field measurements from 3 locations inside Santander.

Moving on to traffic, when talking about the values from the sensors, since most sensors measure in/out rate to the city from the suburbs and overall rest of the world, it seems that rain does not affect the intensity of traffic. In the same week, there is no big irregularity during days when it is raining compared to ones with good weather. This is probably due to the fact that traffic intensity is mainly measured at the main entrances and exits for the city, and not inside the center, where traffic jams could be created more easily due to the weather conditions. Again, long-term measurements show in this case that traffic is half in summer compared to winter. Fig. 15 shows the measurements for traffic intensity against weather conditions during a 2-week time period, while Fig. 16 shows the correlation between the 2 parameters. On average the Pearson correlation coefficient for November and December 2017 was 0.016960 and −0.137907 respectively.

In terms of adding annotations to smart city datasets, through this preliminary evaluation we can claim that there exist a number of scenarios where such annotations could provide missing context to the data. As such an example, anomalies can be detected in traffic intensity, which are not explained by other parameters, e.g., weather. In those cases, citizens can probably add annotations that could explain the situation to a certain extent.

Furthermore, anomaly detection can be used to detect periods where the infrastructure is down, and exclude those values from further analysis that could change average values etc. In general, data anomalies detection can help in identifying hardware malfunctions or failures in specific points. Since there exist clear patterns that are observed throughout the year in all of the datasets we examined here, reporting abnormal values could possibly imply e.g., battery depletion or failure. In the specific case of the e-field measurements, in certain cases, data anomalies were identified that indicated sensors sending measurements with zero values for a large part of the day. This was probably due to the fact that their batteries were beginning to fail, and when recharged through solar or power flowing only for certain hours, the batteries could not last as long as they did previously.

Moreover, irregularities can be identified in the specific case of the parking sensors, where there are periods e.g., when the variability is very small considered to the rest of the days. This could be used to identify days where there are roadworks taking place, which was actually the case in some periods at the center of Santander.

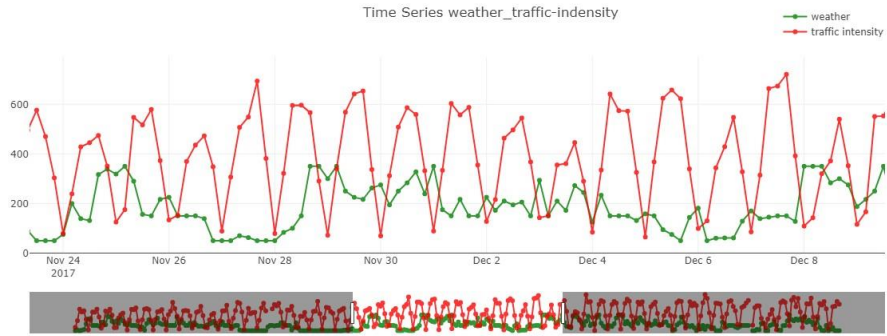

**Fig. 15** Traffic intensity together with weather conditions during 2 weeks in November-December 2017. Even though there are several days in which there was rain (higher values mean worse weather or levels of rain), the values measured for traffic do not seem to be affected much.

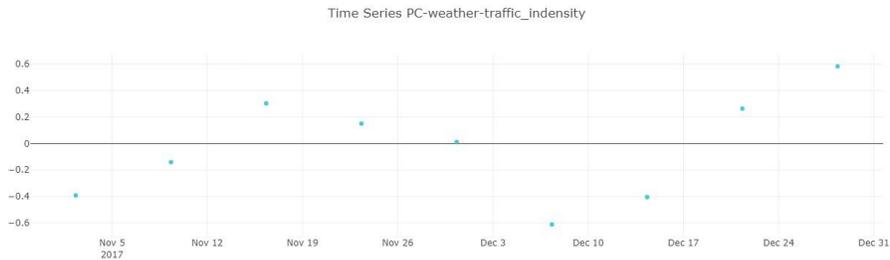

**Fig. 16** Correlation between weather and traffic, showing results that on average there is weak correlation between the two parameters.

## 7 Conclusions and Future Work

Research on smart cities is starting to produce tangible results, introducing new possibilities for city-wide services and applications, as well as accelerating the adoption of new technologies. However, we believe that there is still work to be done with respect to establishing workflows and tools in order to produce results with a more tangible impact on citizens' lives. In this context, one of the directions the research community has recently taken is to combine the existing smart city datasets within machine learning frameworks in more "clever" ways than in the past.

In this article, we presented a solution that validates and enriches data produced by smart city infrastructures. We believe that this kind of processing is critical for IoT sensor data to become something more than simple datasets, i.e., a useful and reliable data source to facilitate the development of future city services. We presented a service capable of automatically detecting erroneous or unexpected sensor data using machine learning algorithms, classifying them and detecting real-world events and situations (in the form of annotation tags) based on provided training data sets.

We have included some examples of analysis of smart city data based on our system, based on parking, noise levels, traffic intensity, mobile network and weather data, produced inside the city centre of Santander. Again, we would like to emphasize that our intention with respect to our evaluation is to showcase the hidden potential of smart city data, which can be revealed given the right tools. Since we are not data scientists, city planners or sociologists, our interpretation of the data presented in this work is bound to be limited by our own scientific background. Having this in mind, our evaluation revealed some interesting findings that could point to novel ways on designing future smart city infrastructure, or how to utilize the combination of data sources that have not been explored yet.

To achieve the next level of smart city data analysis, we believe that citizen participation is of critical importance. We plan on focusing on interactive interfaces that make it easy for users to augment or confirm the automated annotations generated from our system. Moreover, we believe that in order to further increase the participation rate and interest of the citizens, various methods for incentives and gamification should be assessed, either in the context of rewards from the city to the citizens, or in the form of specific benefits/findings that city services could offer to the citizens or local authorities.

**Acknowledgements** This work has been partially supported by the EU research project OrganiCity, under contract H2020-645198.